\documentclass[12pt]{iopart}

\usepackage{graphicx}

\usepackage{iopams}

\usepackage{rotating}

\newtheorem{theorem}{Theorem}

\newtheorem{corollary}{Corollary}

\newtheorem{definition}{Definition}

\newtheorem{notation}{Notation}

\newtheorem{proposition}{Proposition}

\newenvironment{proof}[1][Proof]{\textbf{#1.} }{\ \rule{0.5em}{0.5em}}
\begin{document}

\title[A comment concerning cohomology and invariants of Lie algebras]
{A comment concerning cohomology and invariants of Lie algebras with respect
 to contractions and deformations}

\author{R. Campoamor-Stursberg\dag}

\address{\dag\ Dpto. Geometr\'{\i}a y Topolog\'{\i}a\\Fac. CC. Matem\'aticas\\
Universidad Complutense de Madrid\\Plaza de Ciencias, 3\\E-28040
Madrid, Spain}

\ead{rutwig@mat.ucm.es}

\begin{abstract}
Contrary to the expected behavior, we show the existence of
non-invertible deformations of Lie algebras which can generate
invariants for the coadjoint representation, as well as delete
cohomology with values in the trivial or adjoint module. A
criterion to decide whether a given deformation is invertible or
not is given in dependence of the Poincar\'e polynomial.

\end{abstract}


\maketitle



\section{Introduction}

While contractions were introduced in physics for analyzing the
behavior of systems subjected to some limiting process,
deformations of Lie algebras and their generalizations entered the
theory as an appropriate tool to analyze stability \cite{Vi}. More
recent developments, like the attempt to identify the stable form
of quantum relativistic kinematical algebras, Quantum Field
Theory, expansions of Lie algebras and gauge symmetries or the so
called double Special Relativity also head in that direction
\cite{Fa,Am,Ch,Az,Cha}.

\medskip

Often contractions and deformations are considered as dual
operators. However, this idea, although true in some sense, is not
entirely satisfactory. It is elementary to prove that any Lie
algebra contracts onto the abelian algebra of the same dimension.
However, there is no Lie algebra onto which all algebras of the
same dimension deform. Thus, although contractions determine a
distinguished algebra, the abelian, deformations do not. One could
claim that stable\footnote{Also called rigid Lie algebras.}
algebras play the analogue role for deformations as the abelian
for contractions. However, even this assertion is false since a
Lie algebra does not generally deform onto a stable one. In this
sense, deformations add new possibilities that cannot appear in
limiting processes. Among other properties, contractions of Lie
algebras imply various (numerical) relations among invariants of
contracting and contracted Lie algebras, such as dimensions of
cohomology groups or number of generalized Casimir invariants. It
should therefore be expected that deformations imply the reversal
of these relations. This must obviously be true for deformations
that are the ``inverse" of a contraction, but it will be false for
a generic deformation.

\medskip

More specifically, we show that a non-invertible\footnote{Later
this concept will be made more precise.} deformation can behave in
a rather unexpected way. For example, they can generate Casimir
operators and central extensions, which contradict the expected
pattern that deformations make a Lie algebra ``less abelian". This
shows that, although the comparison of invariants of two given Lie
algebras provides some information whether a contraction between
them is possible, for deformations no assertion can be made by
inspection of the corresponding invariants.

\section{Contractions, deformations and cohomology of Lie
algebras}

Let $\frak{g}$ be  a Lie algebra and $\Phi_{t}\in Aut(\frak{g})$ a
family of automorphisms of $\frak{g}$, where $t\in
[1,\infty)$.\footnote{Other authors use the parameter range
$(0,1]$, which is equivalent to this by simply changing the
parameter to $t^{\prime}=1/t$.} For any $X,Y\in\frak{g}$ we define
\begin{equation}
\left[X,Y\right]_{\Phi_{t}}:=\Phi_{t}^{-1}\left[\Phi_{t}(X),\Phi_{t}(Y)\right],
\end{equation}
which obviously are the brackets of the Lie algebra over the
transformed basis. Now suppose that the limit
\begin{equation}
\left[X,Y\right]_{\infty}:=\lim_{t\rightarrow
\infty}\Phi_{t}^{-1}\left[\Phi_{t}(X),\Phi_{t}(Y)\right]
\label{Ko}
\end{equation}
exists for any $X,Y\in\frak{g}$. Then equation (\ref{Ko}) defines
a Lie algebra $\frak{g}^{\prime}$ called the contraction of
$\frak{g}$ (by $\Phi_{t}$), non-trivial if $\frak{g}$ and
$\frak{g}^{\prime}$ are non-isomorphic, and trivial otherwise. A
contraction for which there exists some basis
$\left\{Y_{1},..,Y_{n}\right\}$ such that the contraction matrix
$A_{\Phi}$ is diagonal, i.e., adopts the form
\begin{equation*}
(A_{\Phi})_{ij}= \delta_{ij}t^{n_{j}},\quad n_{j}\in\mathbb{R},
t>0,
\end{equation*}
is  called a generalized In\"on\"u-Wigner contraction \cite{We}.
An important problem in contraction theory, completely solved in
\cite{We}, is to prove that any contraction is equivalent to a
gen. In\"on\"u-Wigner contraction with integer exponents $n_{j}$.
This implies that any contraction can be realized by diagonal
matrices over some bases.

\medskip

Deformations of Lie algebras arise from the problem of studying
the geometric properties of the variety of Lie algebras when
considered as a transformation space. This leads to study the
neighborhood of a given point (Lie algebra) in the variety.
Deformations are performed using cohomology of Lie algebras
\cite{Ni}. A formal deformation $\frak{g}_{t}$ of a Lie algebra
$\frak{g}=(V,\mu)$ is given by the deformed commutator:
\begin{equation*}
\left[X,Y\right]_{t}:=\left[X,Y\right]+\psi_{m}(X,Y)t^{m},
\end{equation*}
where $t$ is a parameter and $\psi_{m}:V\times V\rightarrow V$ is
a skew-symmetric bilinear map. Imposing the Jacobi identity (up to
quadratic order of $t$) to the deformed commutator, it follows
that the expression satisfied by $\psi_{1}$ characterizes it as a
2-cocycle in the second cohomology space\footnote{See apendix A
for definitions and properties of cohomology.}
$H^{2}(\frak{g},\frak{g})$, i.e., it satisfies the constraint
\begin{eqnarray}
\fl
d\psi_{1}(X_{i},X_{j},X_{k}):=\left[X_{i},\psi_{1}(X_{j},X_{k})\right]+\left[X_{k},\psi_{1}(X_{i},X_{j})\right]+
\left[X_{j},\psi_{1}(X_{k},X_{i})\right]+\nonumber \\
\lo +\psi_{1}(X_{i},\left[X_{j},X_{k}\right])
+\psi_{1}(X_{k},\left[X_{i},X_{j}\right])+\psi_{1}(X_{j},\left[X_{k},X_{i}\right])=0.
\end{eqnarray}

The preceding computation shows that in order to define a Lie
algebra, a deformation has to satisfy an integrability condition.

\begin{definition}
Let $\varphi\in H^{2}(\frak{g},\frak{g})$ be a nontrivial cocycle.
It is called integrable if it satisfies the condition
\begin{equation}
\frac{1}{2}\left[\varphi,\varphi\right](X_{i},X_{j},X_{k}):=\sum_{\sigma\in
S_{3}}
\varphi(X_{\sigma(i)},\varphi(X_{\sigma(j)},X_{\sigma(k)})=0,
\end{equation}
for all $X_{i},X_{j},X_{k}$ in $\frak{g}$.
\end{definition}

Therefore, if $\varphi$ is an integrable cocyle, the linear
deformation given by
\begin{equation*}
\left[X,Y\right]_{t}:=\left[X,Y\right]+t \varphi(X,Y)
\end{equation*}
satisfies the Jacobi identity and defines a Lie algebra. In
particular, nullity of $H^{2}(\frak{g},\frak{g})$ implies that any
deformation is isomorphic to $\frak{g}$ \cite{Ni}.\footnote{Such
algebras are called cohomologically rigid or stable.}

As known, classical kinematical Lie algebras may be seen as
deformations of the static Lie algebra \cite{Ba}. At the same
time, they can be interpreted as contractions of the De Sitter
algebras. This suggests that (physically), contractions and
deformations are inverse procedures. Although it is not globally
true, since there are deformations not related to contractions,
any contraction is actually related to a deformation (see e.g.
\cite{We}).

\begin{definition}
A deformation $\frak{g}_{t}$ ($0\leq t\leq 1$) is called of
plateau type if $\frak{g}_{0}\not\simeq \frak{g}_{1}$ and
$\frak{g}_{t}\simeq \frak{g}_{1}$ for all $t\in (0,1]$.
\end{definition}

The problem of which deformations are related to a contraction is
solved in the following result \cite{We}:

\begin{theorem}
For any contraction $\frak{g}\rightsquigarrow \frak{g}^{\prime}$
there exists a plateau deformation $\frak{g}^{\prime}\rightarrow
\frak{g}$ inverse to the contraction. Conversely, for any
deformation of plateau type there exists a contraction inverse to
it.
\end{theorem}

As a consequence, non-invertible deformations are not of plateau
type. This result moreover indicates for which class of Lie
algebras the invertibility of deformations can fail, namely
families of Lie algebras with some parameter that acts as scaling
factor on some generators.

\section{Number of invariants and deformations}

Since any Lie algebra contracts onto the abelian Lie algebra
$nL_{1}$, in some sense contractions of Lie algebras can be
thought of as an ``abelianizing" operator. Among other properties,
a contraction $\frak{g}\rightsquigarrow\frak{g}^{\prime}$
satisfies the inequalities
\begin{equation}
\begin{array}[l]{l}
\dim H^{j}(\frak{g}) \leq \dim
H^{j}(\frak{g}^{\prime}), \\
\dim H^{1}(\frak{g},\frak{g}) < \dim
H^{1}(\frak{g}^{\prime},\frak{g}^{\prime}),\\
\dim H^{j}(\frak{g},\frak{g}) \leq \dim
H^{j}(\frak{g}^{\prime},\frak{g}^{\prime}), j\neq 1
\end{array} \label{KF}
\end{equation}
for any $j\geq 0$.\footnote{These identities seem to have been
used in the literature without proof. For completeness in the
exposition, in appendix B we give a proof of these inequalities
using exterior forms.} This means that any plateau deformation
reverses the preceding inequalities. Another important property
concerns the number $\mathcal{N}(\frak{g})$ of invariants of the
coadjoint representation. Given a basis $\left\{
X_{1},..,X_{n}\right\} $ of $\frak{g}$ and the structure tensor
$\left\{ C_{ij}^{k}\right\} $, then $\frak{g}$ can be realized in
the space $C^{\infty }\left( \frak{g}^{\ast }\right) $ by means of
the differential operators:
\begin{equation}
\widehat{X}_{i}=C_{ij}^{k}x_{k}\frac{\partial }{\partial x_{j}},
\label{Rep1}
\end{equation}
where $\left[ X_{i},X_{j}\right] =C_{ij}^{k}X_{k}$ \ $\left( 1\leq
i<j\leq n\right) $ and $\left\{ x_{1},..,x_{n}\right\}$ is a dual
basis of $\left\{X_{1},..,X_{n}\right\} $. The invariants of
$\frak{g}$ (in particular, Casimir operators) are the solutions of
the system of partial differential equations:
\begin{equation}
\widehat{X}_{i}F=0,\quad 1\leq i\leq n.  \label{sys}
\end{equation}
The number $\mathcal{N}(\frak{g})$ of functionally independent
solutions is obtained from the classical criteria for differential
equations, and equals:
\begin{equation}
\mathcal{N}(\frak{g}):=\dim \,\frak{g}- {\rm rank}\left(
C_{ij}^{k}x_{k}\right), \label{BB}
\end{equation}
where $A(\frak{g}):=\left(C_{ij}^{k}x_{k}\right)$ is the matrix
associated to the commutator table of $\frak{g}$ over the given
basis. It is known (see e.g. \cite{C24}) that for a contraction
$\frak{g}\rightsquigarrow \frak{g}^{\prime}$ of Lie algebras, the
following inequality must be satisfied
\begin{equation}
\mathcal{N}\left( \frak{g}\right)  \leq\mathcal{N}\left(
\frak{g}^{\prime }\right). \label{KB}
\end{equation}

That is, contractions may generate additional independent
invariants for the coadjoint representation. By theorem 1, any
deformation of plateau type reverses the preceding inequality. It
could therefore be expected that for a deformation
$\frak{g}^{\prime }\rightarrowtail\frak{g}$, even if it not of
plateau type, the following
inequality is satisfied%
\begin{equation}
\mathcal{N}\left(  \frak{g}\right)  \leq\mathcal{N}\left(
\frak{g}^{\prime }\right).
\end{equation}
This agrees with the geometric idea that deformations add more
components to the structure tensor, and therefore can increase the
rank of $A(\frak{g})$. However, we will point out that the latter
equation is generally false, which shows that, in general, there
is no apparent relation between the number of invariants of a Lie
algebra and a deformation which is not of plateau type.

\medskip

To this extent, let
$\mathcal{F}=\left\{\frak{g}_{\alpha}=\frak{sl}\left(
2,\mathbb{R}\right) \overrightarrow
{\oplus}_{2D_{\frac{1}{2}}\oplus D_{0}}A_{5,7}^{1,\alpha,\alpha},
-1\leq \alpha\leq 1\right\}$ be the eight
dimensional family of Lie algebras given by the brackets%
\begin{equation*}
\begin{array}
[l]{llll}%
\left[  X_{1},X_{2}\right]  =2X_{2}, & \left[  X_{1},X_{3}\right]  =-2X_{3}, &
\left[  X_{2},X_{3}\right]  =X_{1}, & \left[  X_{1},X_{4}\right]  =X_{4},\\
\left[  X_{1},X_{5}\right]  =-X_{5}, & \left[  X_{1},X_{6}\right]  =X_{6}, &
\left[  X_{1},X_{7}\right]  =-X_{7}, & \left[  X_{2},X_{5}\right]  =X_{4},\\
\left[  X_{2},X_{7}\right]  =X_{6},\  & \left[
X_{3},X_{4}\right]=X_{5}, & \left[  X_{3},X_{6}\right]  =X_{7}, & \left[  X_{4},X_{8}\right]  =X_{4},\\
\left[  X_{5},X_{8}\right]  =X_{5}, & \left[  X_{6},X_{8}\right]
=\alpha X_{6}, & \left[  X_{7},X_{8}\right]  =\alpha X_{7}. &
\end{array}
\end{equation*}
For the parameter range $-1\leq \alpha\leq 1$, these algebras are
pairwise non-isomorphic \cite{Tu}. The parameter $\alpha$
describes the action of a torus on the maximal nilpotent ideal of
$\frak{g}_{\alpha}$. Using (\ref{BB}) it is straightforward to
verify that
\begin{equation}
\mathcal{N}(\frak{g}_{\alpha}) =\left\{
\begin{array}
[c]{cc}%
0, & \alpha\neq -1\\
2, & \alpha=-1
\end{array}
\right.  .\label{IZ}
\end{equation}
Thus $\frak{g}_{-1}$ is a singular element in the family, and
deserves a more detailed analysis.

\begin{proposition}
$\frak{g}_{-1}$ is the only member of the family $\mathcal{F}$
that is a contraction of a semisimple Lie algebra.
\end{proposition}

\begin{proof}
Since in dimension 8 the only (real) semisimple Lie algebras are
$\frak{su}\left(  3\right)  $, $\frak{su}\left(  2,1\right)  $ and
$\frak{sl}\left(  3,\mathbb{R}\right)  $, if $\frak{g}_{\beta}$ is
a contraction of these for some $\beta\in\left[  -1,1\right]  $,
then it must satisfy $\mathcal{N}\left(  \frak{g}_{\beta}\right)
\geq 2$ by (\ref{KB}). Therefore only $\frak{g}_{-1}$ can appear
as a contraction. In order to obtain it, we analyze the cohomology
group $H^{2}\left(  \frak{g}_{-1},\frak{g}_{-1}\right)  $ and the
existence of invertible (i.e., of plateau type) deformations. A
routine buy tedious computation shows that  $\dim H^{2}\left(
\frak{g}_{-1},\frak{g}_{-1}\right)  =2$, generated by the cocycle
classes of
\begin{eqnarray*}
\begin{tabular}
[c]{ll}%
$\varphi_{1}\left(  X_{4},X_{6}\right)  =X_{2},$ & $\varphi_{1}\left(
X_{4},X_{7}\right)  =-\frac{1}{2}X_{1}+\frac{3}{2}X_{8},$\\
$\varphi_{1}\left(  X_{5},X_{6}\right)  =-\frac{1}{2}X_{1}-\frac{3}{2}X_{8},$%
& $\varphi_{1}\left(  X_{5},X_{7}\right)  =-X_{3}.$%
\end{tabular}
\\
\begin{tabular}
[c]{ll}%
$\varphi_{2}\left(  X_{6},X_{8}\right)  =X_{6},$ & $\varphi_{2}\left(
X_{7},X_{8}\right)  =X_{7}.$%
\end{tabular}
\end{eqnarray*}
Now consider $\frak{g}_{-1}\left(  \varepsilon_{1},\varepsilon_{2}\right)
=\frak{g}_{-1}+\varepsilon_{1}\varphi_{1}+\varepsilon_{2}\varphi_{2}$ with
bracket operation%
\begin{equation*}
\left(  \frak{g}_{-1}+\varepsilon_{1}\varphi_{1}+\varepsilon_{2}\varphi
_{2}\right)  \left(  X,Y\right)  :=\left[  X,Y\right]  +\varepsilon_{1}%
\varphi_{1}\left(  X,Y\right)  +\varepsilon_{2}\varphi_{2}\left(  X,Y\right)
.
\end{equation*}
It is straightforward to verify that the preceding bracket
satisfies the Jacobi condition if and only if\footnote{In this
case the obstruction is
indeed in the third cohomology group $H^{3}\left(  \frak{g}_{-1},\frak{g}%
_{-1}\right)  $, which has dimension 2. See Table 1.}
\begin{equation*}
\varepsilon_{1}\varepsilon_{2} =0.
\end{equation*}
We therefore obtain two types of linear deformations:
$\frak{g}_{-1}\left( \varepsilon_{1}\right)
:=\frak{g}_{-1}+\varepsilon_{1}\varphi_{1}$ and
$\frak{g}_{-1}\left(  \varepsilon_{2}\right)
:=\frak{g}_{-1}+\varepsilon _{2}\varphi_{2}$. Both deformations
are integrable and define a Lie algebra. It follows at once that
$\left[  \frak{g}_{-1}\left(  \varepsilon_{1}\right)
,\frak{g}_{-1}\left(  \varepsilon_{1}\right)  \right]
=\frak{g}_{-1}\left( \varepsilon_{1}\right)  $, thus the deformed
algebra is perfect. In order to prove that $\frak{g}_{-1}\left(
\varepsilon_{1}\right)  $ is semisimple for all values of
$\varepsilon_{1}\neq0$, we compute the Killing metric tensor
$\kappa$. Over the ordered basis $\left\{ X_{1},..,X_{8}\right\}
$, the matrix of $\kappa$ is given by
\begin{equation*}
A_{\kappa}=\left(
\begin{array}
[c]{cccccccc}%
12 & 0 & 0 & 0 & 0 & 0 & 0 & 0\\
0 & 0 & 6 & 0 & 0 & 0 & 0 & 0\\
0 & 6 & 0 & 0 & 0 & 0 & 0 & 0\\
0 & 0 & 0 & 0 & 0 & 0 & -6\varepsilon_{1} & 0\\
0 & 0 & 0 & 0 & 0 & 6\varepsilon & 0 & 0\\
0 & 0 & 0 & 0 & 6\varepsilon_{1} & 0 & 0 & 0\\
0 & 0 & 0 & -6\varepsilon_{1} & 0 & 0 & 0 & 0\\
0 &  & 0 & 0 & 0 & 0 & 0 & 4
\end{array}
\right)  .
\end{equation*}
For $\varepsilon_{1}\neq0$ we have $\det\left(  A_{\kappa}\right)
=-2^{10}3^{7}\varepsilon_{1}^{4}\neq0$, thus $\frak{g}_{-1}\left(
\varepsilon_{1}\right)  $ is semisimple. The eigenvalues of
$A_{\kappa}$ are
\begin{equation*}
Sp(A_{\kappa})=\left\{
4,6,-6,12,6\varepsilon_{1},6\varepsilon_{1},-6\varepsilon
_{1},-6\varepsilon_{1}\right\}  ,
\end{equation*}
therefore for any $\varepsilon_{1}\neq0$ the signature of the matrix is
\begin{equation*}
\sigma=2,
\end{equation*}
proving that $\frak{g}_{-1}\left(  \varepsilon_{1}\right)  $ is
isomorphic to the real normal form $\frak{sl}\left(
3,\mathbb{R}\right)  $ \cite{Co}. Moreover, it follows from this
proof that $\frak{g}_{-1}(\epsilon_{1})$ is a deformation of
plateau type, thus we obtain the contraction
$\frak{sl}(3,\mathbb{R})\rightsquigarrow \frak{g}_{1}$ by
inversion.
\end{proof}

\bigskip

\begin{corollary}
Any element $\frak{g}_{\alpha}$ of the family $\mathcal{F}$ can be
obtained as a linear deformation of $\frak{g}_{-1}$. Additionally,
no algebra $\frak{g}_{\alpha}$ contracts nontrivially onto
$\frak{g}_{-1}$.
\end{corollary}

\begin{proof}
It suffices to consider $0<\varepsilon_{2}\leq2$. Then
$\frak{g}_{-1}\left( \varepsilon_{2}\right)
=\frak{g}_{-1}+\varepsilon_{2}\varphi_{2}$ is isomorphic to
$\frak{g}_{\varepsilon_{2}-1}$ and we have $-1<\varepsilon
_{2}-1\leq1$, thus any member of the family can be reached. To
show that no member of the family contracts onto $\frak{g}_{-1}$,
it suffices to observe
that the dimensions of the derivation algebra are:%
\begin{equation*}
\dim Der\left(  \frak{g}_{\alpha}\right)  =\left\{
\begin{array}
[c]{cc}%
9, & \alpha\neq1\\
11, & \alpha=1
\end{array}
\right.  .
\end{equation*}
Since for any nontrivial contraction $\frak{g\rightsquigarrow
g}^{\prime}$ the strict inequality $\dim Der\left( \frak{g}\right)
<\dim Der\left( \frak{g}^{\prime}\right)  $ \ must be satisfied,
no contraction up to the trivial one is possible. In particular,
this shows that no deformation $\frak{g}_{-1}(\epsilon_{2})$ is of
plateau type.
\end{proof}

\bigskip

\begin{proposition}
For any $-1<\alpha\leq1$ the Lie algebra $\frak{g}_{-1}$ is a
linear deformation of $\frak{g}_{\alpha}$. Moreover the
deformation generates two non-constant invariants for the
coadjoint representation.
\end{proposition}

\begin{proof}
Computing the adjoint cohomology groups for the family
$\mathcal{F}$ (see Table 1) it follows that for any value
$-1<\alpha\leq1$ the Lie algebra $\frak{g}_{\alpha}$ admits the
nontrivial cocycle $\varphi$ defined by
\begin{equation*}
\varphi\left(  X_{6},X_{8}\right)  =X_{6},\;\varphi\left(
X_{7},X_{8}\right) =X_{8}.
\end{equation*}
For $\alpha=\pm1$, this cocycle actually generates the cohomology
space $H^{2}\left( \frak{g}_{\alpha},\frak{g}_{a}\right)  $. Now,
for any  $-1<\alpha\leq1$ the third cohomology group vanishes,
i.e., $H^{3}\left(  \frak{g}_{\alpha },\frak{g}_{\alpha}\right)
=0$, which implies that the linear deformation $\frak{g}_{\alpha
}\left( \varepsilon\right) :=\frak{g}_{\alpha}+\varepsilon\varphi$
is integrable and defines a non-isomorphic Lie algebra. Since for
any value of $\epsilon$ such that $-1\leq \alpha+\epsilon\leq 1$
holds the deformed algebra is nonisomorphic, these deformations
are never of plateau type, showing that no contractions among the
family elements exist. Now, choosing $\epsilon$ such that
$\alpha+\varepsilon=-1$, the Lie algebra $\frak{g}_{\alpha}\left(
-1-\alpha\right)  $ has the same commutators as $\frak{g}_{-1}$,
thus is isomorphic to it. This proves the first assertion. By
(\ref{IZ}), $\frak{g}_{\alpha}$ has no invariants for $\alpha\neq
-1$, thus the deformation decreases the rank of $A(\frak{g})$ and
generates two invariants for the coadjoint
representation.\footnote{These two invariants are actually Casimir
operators, and can be obtained by contraction of the quadratic and
cubic invariants of $\frak{sl}(3,\mathbb{R})$.}
\end{proof}

As a consequence, the deformation decreases the rank of
$A(\frak{g_{\alpha}})$. This means geometrically that the generic
rank of an exterior form in the space spanned by the Maurer-Cartan
forms is reduced by the deformation. This fact is of interest for
representations, since it indicates the possibility that
deformations introduce additional internal labels to describe
basis states of a representation \cite{Ra,C43,C57}.\newline This
result implies the general falseness of the intuitive idea that a
contraction of Lie algebras ``abelianizes" it. The preceding
result show that there exist deformations that delete various
brackets and make the deformed algebra more ``abelian".

\section{Central extensions and cohomology with trivial
coefficients}

It follows from (\ref{KF}) that contractions of Lie algebras can
generate cohomology. The physically most useful situation is that
a trivial central extension $\frak{g}\otimes \mathbb{R}$ of an
algebra $\frak{g}$ leads to a non-trivial central extension
$\widehat{\frak{g}^{\prime}}$ of a contraction
$\frak{g}^{\prime}$. This happens for example for the Poincar\'e
and Galilei algebras, as well as for other kinematical algebras
\cite{Az,Ba}. Obviously deformations of plateau type reverse the
inequalities in (\ref{KF}). We illustrate in this section that
skipping the assumption of plateau type, no assertions can be made
in general on the behavior of the Betti numbers
$b_{i}(\frak{g})=\dim H^{i}(\frak{g}_{\alpha})$ by deformations.

\begin{proposition}
Let $-1\leq \alpha\leq 1$ following relations hold:
\begin{enumerate}

\item For $\alpha\neq-1,0$, the deformations
$\frak{g}_{-1}{\longrightarrow}\frak{g}_{\alpha}$ decrease $b_{2}$
by one unity.

\item For $\alpha\neq-1,0$, the deformations
$\frak{g}_{\alpha}{\longrightarrow}\frak{g}_{\beta}$
($\beta=0,-1$) increase $b_{2}$ by one unity.

\item The deformation $\frak{g}_{-1}{\longrightarrow}\frak{g}_{0}$
preserves $b_{2}$.

\end{enumerate}
\end{proposition}

This implies that a general deformation can also create central
extensions. By (\ref{KF}), any such deformation is not of plateau
type. The proof of this result follows at once by the preceding
results and the dimensions of the cohomology groups
\begin{eqnarray}
H^{2}(\frak{g}_{\alpha})=0,\quad \alpha\neq -1,0, \nonumber\\
\dim H^{2}(\frak{g}_{\alpha})=1, \alpha=-1,0.
\end{eqnarray}

This result has a interesting consequence that allows to determine
whether a given deformation can be of plateau type. Recall that
for any Lie algebra the Poincar\'e polynomial is defined as
\begin{equation}
P_{T}(\frak{g})=1+ \sum_{i=1}^{\dim\frak{g}} b_{i}(\frak{g})T^{i}.
\end{equation}

\begin{proposition}
Let $\frak{g}^{\prime}\longrightarrow \frak{g}$ be a nontrivial
deformation. If the polynomial
$P_{T}(\frak{g}^{\prime})-P_{T}(\frak{g})$ has negative
coefficients, then the deformation cannot be of plateau type.
\end{proposition}

\begin{proof}
If the deformation is of plateau type, then it can be reversed to
a contraction $\frak{g}\rightsquigarrow \frak{g}^{\prime}$ by
theorem 1. By (\ref{KF}) we have $b_{i}(\frak{g})\leq
b_{i}(\frak{g}^{\prime})$ for all $j$. Therefore
\begin{equation}
P_{T}(\frak{g}^{\prime})-P_{T}(\frak{g})=\sum_{i=1}^{\dim\frak{g}}
\left(b_{i}(\frak{g}^{\prime})-b_{i}(\frak{g})\right) T^{i}
\end{equation}
is a polynomial with non-negative coefficients.
\end{proof}

We observe that this result can also be applied as a criterion to
analyze the existence of contractions.

\section{Conclusions}

We have seen that in general there is no possible comparison
between the cohomology and the number of invariants for a
deformation of Lie algebras, up to a special class (plateau type),
which correspond exactly to invertible contractions. Specifically
we have exhibited a family of Lie algebras which, with two
exceptions, does only admit non-invertible deformations. These
deformations can create invariants for the coadjoint
representation, as well as deleting cohomology, contrary to the
expected pattern for deformations. Additionally, these algebras do
not deform onto a stable algebra, with the exception of
$\frak{g}_{-1}$, thus constitute a singular class of algebras in
the variety of Lie algebra laws \cite{Ni}. They moreover provide a
certain geometrical insight for the non-existence of contractions,
since brackets deleted by deformations cannot be recovered by a
limiting process.

\smallskip

Geometrically, the notion of non-plateau type deformations adds a
new perspective to the non-stability of a system. While plateau
deformations give rise to a contraction, thus offering the
transition from one system to the other, non-plateau type
deformations imply not only instability (like all deformations),
but also non-reversibility of processes when subjected to
infinitesimal changes. Consider for instance the Bianchi
VII$_{h}$-type algebras (including VII$_{0}$ type)
\begin{equation}
\left[X_{1},X_{2}\right]= hX_{2}+X_{3},\quad
\left[X_{1},X_{3}\right]=-X_{2}+ hX_{3}. \label{BVII}
\end{equation}
Making the appropriate changes of coordinates, the Bianchi metrics
can be reduced to the following form
\begin{eqnarray}
\fl ds^{2}=e^{-2hx_{3}}\left(
a^{2}(t)(\cos{x_{3}}dx_{1}+\sin{x_{3}}dx_{2})^{2}
+b^{2}(t)(\sin{x_{3}}dx_{1}-\cos{x_{3}}dx_{2})^2\right)\nonumber \\
+c^{2}(t)dx_{3}^{2}-dt^{2}.
\end{eqnarray}
We observe that $\lim_{h\rightarrow 0}ds^{2}$ gives the metric
\begin{equation}
\fl ds^{2}=a^{2}(t)(\cos{x_{3}}dx_{1}+\sin{x_{3}}dx_{2})^{2}
+b^{2}(t)(\sin{x_{3}}dx_{1}-\cos{x_{3}}dx_{2})^2+c^{2}(t)dx_{3}^{2}-dt^{2},
\end{equation}
which actually coincides with the Bianchi metric for type
VII$_{0}$. That is, the metrics of the corresponding models are
related by a limiting process, although for no value $h\neq 0$
(\ref{BVII}) contracts onto VII$_{0}$. However, the latter deforms
to any Bianchi VII$_{h}$ type by means of a deformation of
non-plateau type. Thus introducing an infinitesimal parameter in
VII$_{0}$ leads to a non-equivalent model with quite different
physical properties \cite{Col}. One of these is that this
infinitesimal change in the parameter connects the flat and open
Friedmann-Robertson-Walker models \cite{He}, and that there is no
possibility of recovering the flat model from the open one.

The existence of deformations of non-plateau type could therefore
be of some use in physical models depending on some parameters, in
order to analyze their sensitiveness to small changes (i.e.
perturbations) of the model. These changes could be interpreted as
an additional degree of instability of systems, opposed to stable
systems associated to rigid (e.g. semisimple) Lie algebras.

\begin{table}
\begin{indent}
\caption{Adjoint cohomology of the family $\mathcal{F}$}
\begin{tabular}
[c]{|c|c|c|c|c|}\hline $\alpha$ & $\dim H^{3}\left(
\frak{g}_{\alpha},\frak{g}_{\alpha}\right)  $ & $\dim H^{2}\left(
\frak{g}_{\alpha},\frak{g}_{\alpha}\right)  $ & $\dim Der\left(
\frak{g}_{\alpha}\right)  $ & $\mathcal{N}\left(  \frak{g}_{\alpha
}\right)  $\\\hline
$-1$ & $2$ & $2$ & $9$ & $2$\\
$\alpha\neq\pm1$ & $0$ & $1$ & $9$ & $0$\\
$1$ & $0$ & $3$ & $11$ & $0$\\\hline
\end{tabular}
\end{indent}
\end{table}

\section*{Acknowledgement}
The authors wishes to express his gratitude to J. A. de
Azc\'arraga for reference \cite{Az} and useful comments.

\newpage

\appendix

\section{Cohomology of Lie algebras}

Let $\frak{g}$ be a Lie algebra and $V$ a representation of
$\frak{g}$. A $p$-dimensional cochain of $\frak{g}$ (with values
in $V$) is a $p$-linear skew-symmetric mapping
\begin{equation}
\Phi:\frak{g}\times.^{(p)}.\times\frak{g}\longrightarrow V.
\end{equation}
A $0$-cochain is by definition a constant function. We denote by
$C^{p}\left(  \frak{g},V \right)=Hom(\bigwedge^{p}\frak{g},V) $
the space of $p$-cochains. We can provide $C^{p}\left(
\frak{g},V\right)$ with the structure of a $\frak{g}$-module
structure by putting
\begin{equation}
\fl \left(  X\Phi\right)  \left(  X_{1},...,X_{p}\right)  = X.
\Phi\left( X_{1},...,X_{p}\right)-\sum_{1\leq i\leq p}\Phi\left(
X_{1},...,\left[  X,X_{i}\right]  ,...,X_{p}\right)
\end{equation}
for all $\ X_{1},...,X_{p}\ \in\frak{g}$. The coboundary operator
$d_{p}:C^{p}\left(  \frak{g},V \right)  \longrightarrow C^{p+1
}\left(  \frak{g},V \right)$ is defined by
\begin{eqnarray}
\fl d_{p} \Phi\left(  X_{1},...,X_{p+1}\right)  =\sum_{1\leq s\leq
p+1}\left( -1\right)  ^{s+1}\left(  X_{s}.\Phi\right)  \left(
X_{1},...,{\widehat{X}}_{s},...,X_{p+1}\right)  +\nonumber \\
\lo +\sum_{1\leq s\leq t\leq p+1}\left(  -1\right)
^{s+t}\Phi\left( \left[
X_{s},X_{t}\right],X_{1},...,{\widehat{X}_{s}},...,
{}{\widehat{X}}_{t,...},X_{p+1}\right)
\end{eqnarray}

By this definition, $ d_{p}\left(  C^{p}\left(  \frak{g}%
,\frak{g}\right)  \right)  \subset C^{p+1}\left(
\frak{g},\frak{g}\right)$, and it can be verified  that
$d_{p}\circ d_{p}=0$ for all $p$. The space of $p$-cocycles is
defined as $Z^{p}\left(  \frak{g},V\right)  =\ker d_{p}$, and
coboundaries by $B^{p}\left(  \frak{g},V\right)  ={\rm Im} d_{p}$.
The $p^{th}$-cohomology space with values in $V$ is then defined by
\begin{equation}
H^{p}\left(  \frak{g},V\right)  =Z^{p}\left(  \frak{g}%
,V\right)  /B^{p}\left(  \frak{g},V\right).
\end{equation}
In particular, for any $p\geq$ we have the following identity:
\begin{equation}
\fl \dim B^{p+1}(\frak{g},V)=\dim C^{p}(\frak{g},V)-\dim
Z^{p}(\frak{g},V)=\dim V \left(\begin{array}{c}
\dim \frak{g} \\
p
\end{array}\right)- \dim Z^{p}(\frak{g},V).\label{DF}
\end{equation}

\begin{notation}
For the trivial module $V=\mathbb{R}$ the notation for the
cohomology spaces is simply $H^{p}\left(  \frak{g}\right)$.
\end{notation}

We recall the interpretation of some cohomology groups of low
order (see e.g. \cite{Az2} for further details):

\begin{enumerate}

\item $H^{0}\left( \frak{g},\frak{g}\right)=Z(\frak{g})$ is the
centre of $\frak{g}$.

\item $H^{1}\left(  \frak{g},\frak{g}\right)=Der(\frak{g})$ is the
algebra of derivations.

\item $\dim
H^{1}\left(\frak{g}\right)=$codim$_{\frak{g}}\left[\frak{g},\frak{g}\right]$.

\item $H^{2}\left(\frak{g}\right)$ is identified with the
isomorphism classes of one dimensional central extensions of
$\frak{g}$.
\end{enumerate}

\section{Proof of formula (\ref{KF})}

\begin{proposition}
Let $\frak{g}\rightsquigarrow \frak{g}^{\prime }$ be a
contraction. Then following inequalities hold:

\begin{enumerate}
\item  $\dim H^{k}\left( \frak{g},\mathbb{R}\right) \leq \dim
H^{k}\left( \frak{g}^{\prime },\mathbb{R}\right) ,\;j\geq 0$

\item  $\dim H^{k}\left( \frak{g},\frak{g}\right) \leq \dim
H^{k}\left( \frak{g}^{\prime },\frak{g}\right) ,\;j\geq 0.$
\end{enumerate}
\end{proposition}

\begin{proof}
We prove the result for the De Rham cohomology, the argument being
analogous for the adjoint cohomology. By Theorem 1, there exists a
basis $\left\{ X_{1},..,X_{n}\right\} $ of $\frak{g}$ over which
the contraction is given by the diagonal matrix $T\left(
\varepsilon \right) _{ij}=\delta _{i}^{j}\varepsilon ^{n_{j}}$,
with $n_{j}\in \mathbb{Z}$. We compute the cohomology over the
transformed basis $\left\{ X_{i}^{\prime }=\varepsilon
^{n_{i}}X_{i}\right\} $. A basis of $C^{p}\left(
\frak{g},\mathbb{R}\right) $ is clearly given by the elementary
cochains
\begin{equation}
\varphi _{i_{1}...i_{p}}\left( X_{j_{1}}^{\prime
},..,X_{j_{n}}^{\prime }\right) =\delta _{i_{1}}^{j_{1}}...\delta
_{i_{n}}^{j_{n}}.
\end{equation}
Taking the dual basis $\left\{ \omega _{1},..,\omega _{n}\right\} $ of $%
\left\{ X_{1}^{\prime },..,X_{n}^{\prime }\right\} $, the cocycle
can be identified with the exterior form
\begin{equation}
\varphi _{i_{1}...i_{p}}=\omega _{i_{1}}\wedge ...\wedge \omega
_{i_{p}}.
\end{equation}
Thus an arbitrary $p$-cochain $\varphi $ is given by a
$\mathbb{R}$-linear combination
\begin{equation*}
\varphi =\alpha ^{i_{1}..,i_{p}}\omega _{i_{1}}\wedge ...\wedge
\omega _{i_{p}}.
\end{equation*}
The cocycle condition implies the nullity by the coboundary
operator
\begin{equation}
d\varphi =\alpha ^{i_{1}..i_{p}}d\left( \omega _{i_{1}}\wedge
...\wedge \omega _{i_{p}}\right) =0.  \label{S2}
\end{equation}
Since we are working on the transformed basis, the Maurer-Cartan
equations of $\frak{g}$ have the form:
\begin{equation}
d\omega _{k}=\varepsilon ^{n_{i}+n_{j}-n_{k}}C_{ij}^{k}\omega
_{i}\wedge \omega _{j}.
\end{equation}
Developing (\ref{S2}) we obtain
\begin{equation}
\fl \left( -1\right) ^{i_{l}-1}\alpha ^{i_{1}...i_{p}}\varepsilon
^{n_{j_{l}}+n_{k_{k}}-n_{i_{l}}}C_{j_{l}k_{l}}^{i_{l}}\omega
_{i_{1}}\wedge ..\wedge \omega _{i_{l-1}}\wedge \omega
_{j_{l}}\wedge \omega _{k_{l}}\wedge\omega_{i_{l+1}}\wedge
..\wedge \omega _{i_{p}}=0. \label{S3}
\end{equation}
The latter equation is nothing but a linear system in the $\alpha
^{i_{1}..,i_{p}}$ whose coefficients are of the type $\varepsilon
^{n_{j_{l}}+n_{k_{k}}-n_{i_{l}}}C_{j_{l}k_{l}}^{i_{l}}$. These
coefficients depend both on the structure tensor and the
contraction parameter $\epsilon$. The coefficient matrix of
(\ref{S3}) will be denoted by $A\left( \varepsilon ^{n_{i}}\right)
$. It follows that $Z^{p}\left( \frak{g},\mathbb{R}\right) $ is
given by the kernel of this system, thus
\begin{equation}
\dim Z^{p}\left( \frak{g},V \right) =\left(
\begin{array}{c}
\dim \frak{g} \\
p
\end{array}
\right) -{\rm rank}\;A\left( \varepsilon ^{n_{i}}\right) .
\end{equation}
Observe that the coefficient matrix depends on $\varepsilon
^{n_{i}}$. Thus taking the limit we have
\begin{equation}
{\rm rank}\,A\left( \varepsilon ^{n_{i}}\right) \geq {\rm
rank}\;\left( \lim_{\in \rightarrow 0}\;A\left( \varepsilon
^{n_{i}}\right) \right) .
\end{equation}
But $B:=\left( \lim_{\in \rightarrow 0}\;A\left( \varepsilon
^{n_{i}}\right) \right) $ is the coefficient matrix we obtain
computing the cohomology after the contraction, thus we get
\begin{equation}
\left(
\begin{array}{c}
\dim \frak{g} \\
p
\end{array}
\right) -{\rm rank}\left( \lim_{\in \rightarrow 0} A\left(
\varepsilon ^{n_{i}}\right) \right) =\dim Z^{p}\left(
\frak{g}^{\prime},\mathbb{R}\right).
\end{equation}
This proves that $\dim Z^{p}\left( \frak{g},\mathbb{R}\right) \leq
\dim
Z^{p}\left( \frak{g}^{\prime },\mathbb{R}\right) $ for any $p$. By (\ref{DF}%
) it follows at once that
\begin{equation}
\dim B^{p}\left( \frak{g},\mathbb{R}\right) \geq \dim B^{p}\left( \frak{g}%
^{\prime },\mathbb{R}\right) ,\;p\geq 0.
\end{equation}
Putting together these inequalities we obtain the chain
\begin{eqnarray}
 \dim Z^{p}\left( \frak{g}^{\prime },\mathbb{R}\right) -\dim
B^{p}\left( \frak{g}^{\prime },\mathbb{R}\right) \geq \dim
Z^{p}\left( \frak{g}^{\prime },\mathbb{R}\right) -\dim B^{p}\left(
\frak{g},\mathbb{R}\right)    \nonumber\\
  \geq \dim Z^{p}\left( \frak{g},\mathbb{R}\right) -\dim B^{p}\left( \frak{g},\mathbb{R}%
\right) ,
\end{eqnarray}
that is,
\begin{equation}
\dim H^{p}\left( \frak{g}^{\prime },\mathbb{R}\right) \geq \dim
H^{p}\left( \frak{g},\mathbb{R}\right) .
\end{equation}

\end{proof}

As a direct consequence of this formula, no contraction $\frak{g}$
of a semisimple Lie algebra $\frak{s}$ can have vanishing
cohomology $H^{3}(\frak{g})$.

\end{document}